\documentclass[pdflatex,sn-mathphys-num,iicol]{sn-jnl}

\usepackage{amsmath,amssymb,amsfonts}
\usepackage{graphicx}
\usepackage{booktabs}
\usepackage{siunitx}
\usepackage{xcolor}
\usepackage{xurl}

\newcommand{\mass}{m_\chi}

\newcommand{\poi}{\mu}
\newcommand{\nuis}{\psi}
\newcommand{\given}{\,|\,}

\newcommand{\Prob}{\mathbb{P}}
\newcommand{\Poisson}{\mathrm{Poisson}}
\newcommand{\nubb}{0\nu\beta\beta}

\begin{document}

\title[Silent coverage failures in rare-event searches and a degeneracy index that predicts them]{Silent coverage failures in rare-event searches and a degeneracy index that predicts them}

\author*[1,2]{\fnm{Davide} \sur{Pagano}}

\affil*[1]{\orgdiv{Department of Mechanical and Industrial Engineering},
  \orgname{University of Brescia},
  \orgaddress{\city{Brescia}, \country{Italy}}}

\affil[2]{\orgname{Istituto Nazionale di Fisica Nucleare (INFN)},
  \orgaddress{\city{Pavia}, \country{Italy}}}

\abstract{Searches for new physics in low-background experiments infer a
non-negative signal strength from few events and often report an upper limit.
Nominal frequentist coverage requires both a valid interval construction and
an adequate data model. We study how controlled model departures affect
lower- and upper-endpoint coverage for six interval procedures in an exact
Poisson counting experiment, dark-matter recoil spectra, and a
neutrinoless-double-beta-decay peak search. We introduce the Poisson--Fisher
degeneracy index
$\mathcal I_{\mathrm{PF}}(\delta\nu;\vartheta_0)=(\beta,\gamma)$, which maps
a specified expected-count deformation, after projection onto the complete
fitted tangent space, to $\beta$, the signed fitted signal shift in profiled
standard-error units, and $\gamma$, the Poisson--Fisher norm of the unabsorbed
residual. Locally, the sign of $\beta$ identifies the threatened endpoint,
while $\gamma$ provides a local indication of detectability by the
saturated-Poisson goodness-of-fit test used here at a $5\%$ nominal type-I
error rate. Across the
studied deformations, positive signal-like bias degrades discovery-side
coverage while making upper limits conservative; negative signal bias from
overestimated signal efficiency can make the upper endpoint undercover.
Calibration under the nominal simulator does not protect against
misspecification of that simulator relative to the data-generating process. A
plausible deformation with large $|\beta|$ and small $\gamma$ may therefore
evade diagnosis and should be represented by a nuisance constrained with
auxiliary information or included in a defensible envelope. In the exactly
collinear constrained-nuisance benchmark, modelling the deformation restores
coverage at the evaluated grid points, at a quantifiable cost in interval
sensitivity.}


\maketitle

\section{Introduction}
\label{sec:intro}

Searches for physics beyond the Standard Model often proceed by exclusion.
Direct dark-matter detectors such as LZ~\cite{LZ2024},
XENONnT~\cite{XENONnT2023}, and SuperCDMS~\cite{SuperCDMS2018} seek WIMP-induced
nuclear recoils, whereas $\nubb$ experiments such as GERDA~\cite{GERDA2020},
CUORE~\cite{CUORE2022}, and KamLAND-Zen~\cite{KamLANDZen2023} seek a
monoenergetic peak above a continuous background. In both settings the
parameter of interest is non-negative, signal counts are small, backgrounds
are uncertain, and the result is often an \emph{upper limit}. Its frequentist
interpretation depends on \emph{coverage}: at every fixed true value, a
nominal $90\%$ confidence interval should contain that value in at least
$90\%$ of repeated experiments.

Two features make coverage difficult to guarantee. First, the
\emph{boundary}: the parameter of interest sits at or near the edge of its physical
domain, $\poi \geq 0$, where the regularity conditions underpinning the standard
asymptotic ($\chi^2$, Wilks) theory fail~\cite{Wilks1938,Chernoff1954,Cowan2011},
including in the presence of nuisance parameters~\cite{SelfLiang1987},
and low counts require care to avoid
empty or pathological intervals~\cite{FeldmanCousins1998}. Second, and more insidiously,
\emph{model misspecification}: the likelihood (or simulator) used for inference is never
identical to the process that generated the data. An unmodelled low-energy background
tail, an uncertain dark-matter halo model, or an optimistic detector response
can therefore invalidate classical coverage guarantees.
We call a coverage failure \emph{undetected} (or ``silent'') when the
interval misses its stated frequentist target while the accompanying
goodness-of-fit test has rejection probability close to its nominal
false-positive rate.

For a well-specified low-count model, Feldman--Cousins intervals have
finite-sample coverage by construction~\cite{FeldmanCousins1998};
$\mathrm{CL}_s$ avoids excluding signals to which the experiment has little
sensitivity, at the cost of conservatism~\cite{Read2002,Junk1999}. Asymptotic
profile-likelihood intervals are fast but can miscover at low counts and near
boundaries~\cite{Cowan2011}. Bayesian credible intervals depend on the prior
and need not have frequentist coverage.

When the likelihood is intractable, simulation-based inference
(SBI)~\cite{Cranmer2020} learns from simulator draws. Our benchmark retains an
explicit binned likelihood as a controlled reference for two SBI-era
constructions: LF2I, which wraps an exact or learned statistic in a
simulator-calibrated Neyman construction and audits point-wise
coverage~\cite{Dalmasso2021,Masserano2023,LF2I2024}, and a split-conformal
posterior-score construction, ``conformal-NPE'', related to conformal SBI
methods such as CANVI~\cite{Patel2024}, whose guarantee is marginal over the
chosen prior predictive. Learned posteriors can be inaccurate even under the
simulator~\cite{Hermans2022}, simulator mismatch motivates discrepancy
models~\cite{Cannon2022,Ward2022}, and conformal validity requires
exchangeability~\cite{Tibshirani2019,Barber2023}.

Fisher geometry, nuisance projection, and misspecification asymptotics supply
important ingredients, but they do not by themselves provide a single signed
diagnostic for three linked questions in low-count, boundary-constrained
inference: whether a physically specified model error can bias the fitted
signal, which interval endpoint is threatened, and whether the fitted
residual can reveal the error. We introduce that missing construction and
assess it in controlled rare-event searches. Our main contributions are:
\begin{enumerate}
  \item The \textbf{Poisson--Fisher degeneracy index}
        $\mathcal I_{\mathrm{PF}}=(\beta,\gamma)$, which pairs the signed
        standardized signal bias with the residual left outside the complete
        fitted tangent space, together with a step-by-step procedure for its
        use in an analysis (Section~\ref{sec:results-index}).
  \item A \textbf{controlled-severity protocol} for positive and negative
        deformations that measures lower-endpoint, upper-endpoint, and
        whole-interval coverage separately and tests projection onto the full
        fitted tangent space
        (Sections~\ref{sec:misspec} and~\ref{sec:results-spectral-efficiency}).
  \item A \textbf{coverage--sensitivity remedy}: promoting a signal-like
        deformation to a constrained nuisance restores coverage at the
        evaluated grid points, at a measurable cost in interval sensitivity
        (Section~\ref{sec:results-nuisance}).
  \item An \textbf{empirical assessment} in an exact Poisson counterexample,
        dark-matter recoil spectra, and a $\nubb$ peak search
        (Sections~\ref{sec:results} and~\ref{sec:results-index-validation}).
\end{enumerate}

\section{Statistical setting and analysis design}
\label{sec:setting}

We consider a non-negative parameter of interest ($\poi\geq0$), optional
additional reported physics parameters $\xi$, and nuisance parameters
$\nuis$. We write $\theta=(\poi,\xi)$ for the reported parameter vector and
$\vartheta=(\theta,\nuis)$ for the full fitted parameter vector; components
that are absent in a particular analysis are simply omitted. In a single
region of interest (ROI), with a fixed non-negative background $b$, the
canonical likelihood is
\begin{equation}
  n \sim \Poisson(\poi\, \kappa + b),
  \label{eq:counting}
\end{equation}
where $\kappa$ is the expected signal count at $\poi = 1$ and $b$ the
expected background.
A confidence-region procedure maps an observation to a set
$\mathcal{R}_\alpha(x)\subseteq\Theta$, where $\Theta$ is the space of
reported parameters, and is intended to satisfy, at confidence level
$1-\alpha$, the \emph{coverage} property
\begin{equation}
  \Prob_{x \sim p(\cdot \given \vartheta)}\!\big[\,
  \theta \in \mathcal{R}_\alpha(x) \,\big]
  \;\geq\; 1-\alpha .
  \label{eq:coverage}
\end{equation}
Equation~\eqref{eq:coverage} defines \emph{conditional} or point-wise
coverage at fixed $\vartheta$. For a result reported only in $\poi$, or for
the $\poi$ projection of a joint region, we write
$\mathcal C_\alpha^\poi(x)=\operatorname{proj}_\poi
\mathcal R_\alpha(x)=[L(x),U(x)]$. Marginal coverage instead averages over a
declared parameter distribution $\pi(\vartheta)$:
\begin{equation}
  \Prob_{\vartheta\sim\pi,\; x \sim p(\cdot\given\vartheta)}
  \!\big[\, \theta \in \mathcal{R}_\alpha(x) \,\big]
  \;\geq\; 1-\alpha .
  \label{eq:marginal}
\end{equation}
This is the finite-sample conformal guarantee under exchangeability when the
score is fixed before calibration. It does not imply
Eq.~\eqref{eq:coverage}, which may fail at particular values, especially near
boundaries; we therefore label the two metrics separately.

\paragraph*{Why the boundary is challenging.}
When the remaining parameters are regular and identified, a likelihood-ratio
statistic at $\poi=0$ has the familiar ``half-$\chi^2$'' limit rather than the
usual $\chi^2_1$ law for an interior parameter point~\cite{Chernoff1954,Cowan2011}.
An additional signal-shape parameter $\xi$ may instead be unidentified under
the no-signal hypothesis, as the WIMP mass $\mass$ is when $\poi=0$, and the
statistic then has a different nonstandard distribution. Together with
discreteness and small counts, this motivates exact or calibrated
constructions; the counting study admits an exact Neyman reference.

\paragraph*{Why misspecification is worse.}
Equations~\eqref{eq:coverage}--\eqref{eq:marginal} presuppose the correct
sampling distribution at the true physical parameter value. A deformation
can place the data-generating distribution outside the nominal family or
make it coincide with a nominal distribution at a different parameter value.
In the latter case, the fit can describe the data perfectly while the
interval misses the physical parameter of interest. For example, a uniform
signal-efficiency loss $f$ in Eq.~\eqref{eq:counting} gives
$p^\star(n\mid\poi)=p_0(n\mid(1-f)\poi)$: the nominal family contains the
distribution, but assigns it the wrong signal strength. Neither case
automatically preserves coverage of the physical parameter. The question is
empirical: how fast does the
\emph{effective} coverage,
\begin{equation}
  c_{\mathcal R}(\vartheta;s) \;=\;
  \Prob_{x \sim p^\star_s(\cdot\given\vartheta)}\!\big[\,
  \theta \in \mathcal{R}_\alpha(x) \,\big],
  \label{eq:effective}
\end{equation}
fall below $1-\alpha$ as $p^\star$ departs from the assumed model, and does any
diagnostic track that departure? We measure this region coverage or its
projected-interval analogue, as appropriate.

\subsection{Analysis settings}
\label{sec:simulator}

We model a low-background search in two main settings. Throughout, we
distinguish the \emph{data-generating model}, from which pseudo-data are
sampled, from the \emph{nominal analysis model} used to construct intervals.
Here model misspecification means that these distributions differ at the
same physical parameter value, including the exactly degenerate case
described above. Numerical scales follow generic xenon time-projection chambers
such as LZ and XENONnT~\cite{LZ2024,XENONnT2023}.

\paragraph*{The counting study.}
The first setting is a single energy region of interest described by the
Poisson likelihood in Eq.~\eqref{eq:counting}. The parameter of interest
$\poi\geq0$ is proportional to the WIMP--nucleon cross-section through the
conversion factor $\kappa$, which contains rate, exposure, and efficiency, while
the background $b$ is fixed and known.

\paragraph*{The spectral analysis.}
We extend the single-bin approach to the binned nuclear-recoil energy
spectrum. The WIMP signal has a falling, mass-dependent
shape~\cite{LewinSmith1996} and is passed through an energy-dependent
efficiency, Gaussian resolution, and an analysis threshold. The benchmark
admits an explicit binned-Poisson likelihood, which provides exact reference
statistics for the simulation-based constructions. The inferred parameters
are $(\poi,\mass)$, with $\poi \geq 0$.

\begin{figure*}[t]
  \centering
  \includegraphics[width=\textwidth]{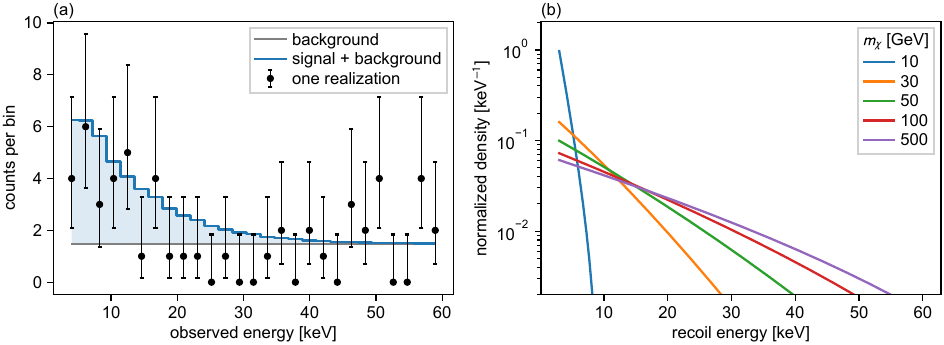}
  \caption{\textbf{Spectral forward model.} (a) Expected signal and
  background at $(\poi,\mass)=(1,\SI{50}{GeV})$ after efficiency, Gaussian
  resolution, and threshold effects; points are one Poisson realization with
  $68\%$ central Poisson (Garwood) intervals~\cite{Garwood1936}.
  (b) Unit-normalized signal spectra for the indicated WIMP masses.}
  \label{fig:level1}
\end{figure*}

\section{Controlled model misspecification}
\label{sec:misspec}

We represent each source of misspecification as a deformation of the nominal
expected-count vector,
\begin{equation}
  \nu^\star(\vartheta;s)=\nu_0(\vartheta)+\delta\nu(\vartheta;s),
  \qquad \delta\nu(\vartheta;0)=0,
  \label{eq:controlled-deformation}
\end{equation}
where $s\in[0,1]$ is the severity and $\vartheta=(\theta,\nuis)$ the full
parameter vector defined above. Pseudo-data
are drawn from $\nu^\star$, but every interval is constructed with the fixed
nominal model $\nu_0$. The severity $s=1$ denotes a declared stress-test
endpoint, not a universal physical scale. This formulation covers additive
rate or shape contamination, signal-model deformation, and detector-response
error; the numerical stress-test endpoints are defined below.

\begin{description}
  \item[Unmodelled background tail.] The analysis assumes a flat/Gaussian background in
    the ROI, but the true background carries an extra low-energy tail --- e.g.\ from a
    residual radioactive contaminant or leakage of surface events. The severity scales
    an exponential shape, normalized over the ROI,
    \[
     f_i\propto\int_{E_i}^{E_{i+1}}e^{-(E-E_{\rm thr})/E_d}\,dE .
    \]
    The counting study uses only its integrated excess, up to six events; the spectral
    study uses $E_d=\SI{1}{keV}$ and up to 40 events
    (Section~\ref{sec:results}).
  \item[Shifted dark-matter halo.] The predicted recoil spectrum depends on
    the galactic velocity distribution, a widely discussed astrophysical
    uncertainty~\cite{Green2017}. As severity grows, the data-generating halo
    parameters interpolate linearly from the nominal Standard Halo Model to
    the deliberately severe alternative
    $v_0:\SI{220}{km\,s^{-1}}\to\SI{420}{km\,s^{-1}}$,
    $v_{\rm Earth}:\SI{232}{km\,s^{-1}}\to\SI{352}{km\,s^{-1}}$, and
    $v_{\rm esc}:\SI{544}{km\,s^{-1}}\to\SI{484}{km\,s^{-1}}$.
    Both the accepted yield and shape may change after thresholding. The
    endpoint is a geometric stress test, not a calibrated astrophysical
    uncertainty envelope.
  \item[Unmodelled $\gamma$ line ($\nubb$ domain).] The analysis assumes a
    flat continuum around the $Q$-value, whereas the true background contains
    a weak peak from a rare or unexpected contaminant. Writing $\Phi$ for the
    standard-normal CDF and $u_i=(E_i-E_\ell)/\sigma_E$, its bin counts are
    \[
     \ell_i(s)=sA_{\max}\big[\Phi(u_{i+1})-\Phi(u_i)\big],
    \]
    with $A_{\max}=6$ and $\sigma_E=\SI{1.2}{keV}$. The line position
    $E_\ell$ relative to the $Q$-value dictates its degree of degeneracy with
    the signal peak (Section~\ref{sec:results-nubb}).
  \item[Overestimated signal efficiency.] The analysis assumes the nominal
    signal efficiency, whereas the data-generating model has a smaller signal
    yield per unit $\poi$, a controlled negative signal bias that tests whether
    upper endpoints become spuriously stringent. In the spectral extension we
    compare a uniform loss with the logistic turn-on
    \[
     \epsilon(E;E_{\rm th})=\frac{\epsilon_\infty}{1+e^{-(E-E_{\rm th})/w}},
    \]
    using $\epsilon_\infty=0.9$, $w=\SI{0.5}{keV}$, and a true threshold
    shifted from \SI{3}{keV} to as much as \SI{9}{keV}. The latter changes both signal yield and shape
    (Sections~\ref{sec:results-edge} and~\ref{sec:results-spectral-efficiency}).
\end{description}

\section{The Poisson--Fisher degeneracy index: damage and detectability}
\label{sec:results-index}

We introduce the Poisson--Fisher degeneracy index as a two-coordinate,
nuisance-aware diagnostic of a candidate local model deformation. It is
intended to screen potential coverage damage and goodness-of-fit
detectability before toy calibration, using only expected bin counts and
local model derivatives.

Choose a reference point $\vartheta_0$ in the fitted model. Let
$\nu_0=\nu_0(\vartheta_0)$ be the corresponding vector of nominal expected
bin counts, and let $\delta\nu=\delta\nu(\vartheta_0;s)$ from
Eq.~\eqref{eq:controlled-deformation} be the change in expected counts under
the candidate data-generating model. Within each locally identifiable fitted
branch, assume $\nu_{0,i}>0$ in every retained bin. For two perturbation
vectors $u$ and $v$, define the Poisson--Fisher inner product
\begin{equation}
 \langle u,v\rangle_W
 =u^{\mathsf T}Wv
 =\sum_i\frac{u_i v_i}{\nu_{0,i}},
 \label{eq:fisher-inner-product}
\end{equation}
with $W=\operatorname{diag}(1/\nu_{0,i})$.
This is the local quadratic metric obtained by expanding the Poisson deviance
around $\nu_0$; it gives greater weight to a fixed absolute deviation in a bin
with a smaller nominal mean. Let $T$ be the local tangent matrix whose column
for parameter $\vartheta_a$ is $\partial\nu/\partial\vartheta_a$, evaluated
at $\vartheta_0$. Its columns include the reported coordinates $\theta$, such
as $\poi$ and $\mass$, and all nuisance parameters $\nuis$, including
background normalizations and shape parameters. Here a \emph{template} is a
vector of expected bin counts, and the columns of $T$ are its local derivative
templates. With the Fisher information $F=T^{\mathsf T}WT$, the parameter
shift that best reproduces $\delta\nu$ to first order, and the part left
outside the fitted tangent space, are
\begin{equation}
\begin{aligned}
  \Delta\hat{\vartheta}&=F^{-1}T^{\mathsf T}W\delta\nu,
  \qquad
  \beta=\frac{\Delta\hat\poi}
  {\sqrt{(F^{-1})_{\poi\poi}}},\\
  \gamma^2&=
  (\delta\nu-T\Delta\hat{\vartheta})^{\mathsf T}
  W(\delta\nu-T\Delta\hat{\vartheta}) .
\end{aligned}
  \label{eq:index}
\end{equation}
We call
$\mathcal I_{\mathrm{PF}}(\delta\nu;\vartheta_0)\equiv(\beta,\gamma)$ the
Poisson--Fisher degeneracy index. The signed coordinate $\beta$ is the
predicted shift in $\poi$, expressed in profiled standard-error units; it
therefore measures the deformation's potential inferential damage. The
non-negative coordinate $\gamma$ is the Poisson--Fisher norm of the residual
after all fitted directions have been profiled and measures the mismatch
available to a compatible residual-based goodness-of-fit test. Because the
projection includes nuisance directions, a deformation parallel to a
background derivative is absorbed mainly by that nuisance rather than by
$\poi$.

If a secondary signal parameter is discrete or unidentified at the boundary,
the projection is evaluated separately in each candidate local tangent space
and the best-fitting branch is used. In the one-template, no-nuisance limit,
with $t=\partial\nu/\partial\poi$, Eq.~\eqref{eq:index} reduces to the signed
matched-filter projection
$\beta=\langle\delta\nu,t\rangle_W/\lVert t\rVert_W$, together with its
orthogonal residual.

Equation~\eqref{eq:index} extends to constrained parameters by treating the
primary and auxiliary measurements jointly. Let $T_{\rm prim}$ be the
Jacobian of the primary expected counts with respect to the full fitted
parameter vector at $\vartheta_0$, and $F_{\rm aux}$ the Fisher information of
an independent, correctly specified auxiliary measurement in the same
coordinates. If the deformation affects only the primary counts,
\[
\begin{aligned}
 F_{\rm tot}&=T_{\rm prim}^{\mathsf T}WT_{\rm prim}+F_{\rm aux},\\
 \Delta\hat{\vartheta}&=F_{\rm tot}^{-1}T_{\rm prim}^{\mathsf T}W\delta\nu,\\
 \beta&=\Delta\hat\poi\big/\sqrt{(F_{\rm tot}^{-1})_{\poi\poi}},
\end{aligned}
\]
and a goodness-of-fit assessment of both measurements sees the joint residual
\[
 \gamma_{\rm tot}^2=\lVert\delta\nu-T_{\rm prim}\Delta\hat{\vartheta}\rVert_W^2
 +\Delta\hat{\vartheta}^{\mathsf T}F_{\rm aux}\Delta\hat{\vartheta},
\]
whose second term is the tension created in the auxiliary measurement.

A background-efficiency loss $f_i$ gives $\delta\nu_i=-f_i b_i$, where $b_i$
is the nominal background expectation. In the one-template, no-nuisance limit,
\[
 \beta=
 -\frac{\sum_i f_i b_i t_i/\nu_{0,i}}
 {\sqrt{\sum_i t_i^2/\nu_{0,i}}}\;\leq\;0
\]
for non-negative signal and background templates. A uniform loss with a
freely fitted background normalization lies along that nuisance tangent and
gives $\beta=\gamma=0$; an energy-dependent $f_i$ projects onto all fitted
background and signal columns of $T$, whose signal coordinate gives $\beta$
and whose residual gives $\gamma$. An auxiliary constraint prevents complete
absorption through $F_{\rm aux}$. If the profiled information on $\poi$
vanishes, $\poi$ is locally non-identifiable and no finite standardized
displacement should be reported.

The Fisher metric and nuisance-tangent projections are established in
information geometry and likelihood theory~\cite{Amari1985,CoxReid1987}.
Huber and White establish pseudo-true displacements and sandwich uncertainty
under misspecification~\cite{Huber1967,White1982}, which do not restore
coverage of a physical parameter that differs from the pseudo-true target,
and the usual
likelihood-ratio $\chi^2$ reference can also fail~\cite{Kent1982}. Bonhomme
and Weidner relate local target-parameter sensitivity and robust confidence
intervals to the local power of a specification
test~\cite{BonhommeWeidner2022}. To our knowledge, none of these works gives
the construction introduced here: a projection of a specified Poisson
expected-count deformation through the complete fitted tangent space that
retains the signed standardized parameter-of-interest displacement, pairs it
with the orthogonal Poisson--Fisher residual, and connects both prospectively
to endpoint-specific coverage vulnerability and residual-test detectability.

The endpoint relations follow from an explicit local-normal approximation.
For a contiguous deformation, regular interior fitted parameters, and
nominal model-based standard error
$\sigma_\poi^2=(F^{-1})_{\poi\poi}$, the quadratic likelihood gives
$\widehat\poi-\poi\ \dot\sim\ \mathcal N(\Delta\widehat\poi,\sigma_\poi^2)$,
so that $\beta=\Delta\widehat\poi/\sigma_\poi$.
Writing $z_q=\Phi^{-1}(q)$ and denoting lower- and upper-endpoint coverage by
$c_L=\Pr[L(X)\leq\poi]$ and $c_U=\Pr[U(X)\geq\poi]$, respectively,
approximating a central interval by
$[\widehat\poi-z_{1-\alpha/2}\sigma_\poi,\,
\widehat\poi+z_{1-\alpha/2}\sigma_\poi]$ therefore yields
\begin{equation}
\begin{aligned}
c_L
&=\Pr(\widehat\poi-\poi\leq z_{1-\alpha/2}\sigma_\poi)
 \simeq\Phi(z_{1-\alpha/2}-\beta),\\
c_U
&=\Pr(\widehat\poi-\poi\geq-z_{1-\alpha/2}\sigma_\poi)
 \simeq\Phi(z_{1-\alpha/2}+\beta).
\end{aligned}
\label{eq:local-endpoint-coverage}
\end{equation}
This is the familiar relation between a standardized bias and one-sided
coverage, and it is not specific to low counts. What the index adds is a
prospective evaluation of $\beta$, before any toy calibration and through the
complete fitted tangent space, whose result can differ substantially from a
single-template estimate (Section~\ref{sec:results-spectral-efficiency}),
paired with the residual $\gamma$ that provides a local indication of the
same deformation's detectability by a compatible goodness-of-fit test.
Under fixed, nonlocal misspecification, the asymptotic covariance is generally
the sandwich covariance $H^{-1}JH^{-1}$, rather than $F^{-1}$. If, in addition,
the true parameter lies on a physical boundary or a secondary parameter is
unidentified under the null, the estimator or test statistic may have a
nonstandard limiting distribution.
Equation~\eqref{eq:local-endpoint-coverage} is consequently a local reference
for sign and severity, not a finite-sample coverage calibration. By contrast,
$\gamma^2$ is the local noncentrality contribution for the quadratic
residual-based goodness-of-fit test used here. Large $|\beta|$ with small
$\gamma$ is the dangerous regime: the interval can move while the fitted
spectrum changes little.

\paragraph*{Using the index in an analysis.}
The index is meant to be computed before any toy calibration, as the first
step of a robustness study:
\begin{enumerate}
  \item \emph{Specify candidate deformations.} Before examining the search
        data, write each plausible model error as a change $\delta\nu$ of the
        expected counts, taken from alternative simulations, calibration
        uncertainties, or control-sample fits, with an amplitude fixed by
        calibration bounds or a declared envelope.
  \item \emph{Compute $(\beta,\gamma)$.} Evaluate Eq.~\eqref{eq:index} from
        the nominal templates and their derivatives with respect to all fitted
        parameters, including nuisances and auxiliary constraints; no
        pseudo-experiments are needed.
  \item \emph{Read the sign of $\beta$.} A positive $\beta$ threatens the
        lower endpoint (false rejection of $\poi=0$), a negative $\beta$ the
        upper endpoint (an over-stringent limit);
        Eq.~\eqref{eq:local-endpoint-coverage} gives the local size.
  \item \emph{Use $(\beta,\gamma)$ to guide model checks and uncertainty treatment.} A deformation with
        appreciable $\gamma$ leaves a residual that a goodness-of-fit test can
        detect: keep a test whose binning and range retain sensitivity to it.
        Detectability alone does not restore coverage; a plausible deformation
        that threatens an endpoint also requires an appropriate uncertainty
        treatment, even when a goodness-of-fit test has power against it.
        For a plausible deformation with large $|\beta|$ and small $\gamma$,
        the goodness-of-fit test may have little power: promote it to a nuisance template,
        $\nu\to\nu+\eta\,\delta\nu$, constrained by auxiliary data or bounded
        by a declared envelope, and report the resulting loss of sensitivity.
  \item \emph{Confirm with toys.} Calibrate the final interval and evaluate
        $c_L$ at $\poi=0$ and $c_U$ at nonzero $\poi$, for deformations of
        both signs.
\end{enumerate}
There are no universal thresholds for a ``large'' $|\beta|$ or a ``small''
$\gamma$: an analysis should derive them from the largest tolerable change in
endpoint coverage and the required model-check power.
Table~\ref{tab:summary} in Section~\ref{sec:results-index-validation} illustrates
steps~2--4 for representative configurations of the deformation families.

\section{Inference methods and numerical protocol}
\label{sec:protocol}

The counting studies use Feldman--Cousins, $\mathrm{CL}_s$, profile-likelihood,
and Bayesian intervals; the spectral studies use profile likelihood, LF2I,
and split-conformal posterior scores. Feldman--Cousins and LF2I use
model-specific Neyman calibration, the baseline profile likelihood is
asymptotic, the
$\mathrm{CL}_s$ construction used here is conservative, Bayesian intervals
control posterior credibility, and the conformal guarantee is
prior-predictive and marginal.
None is automatically valid outside its analysis model.

\paragraph*{Counting constructions.}
Feldman--Cousins uses likelihood-ratio ordering of the Poisson counts.
For the one-sided construction,
$\mathrm{CL}_s(\poi)=\mathrm{CL}_{s+b}(\poi)/\mathrm{CL}_b$, with both
terms given by the corresponding lower-tail Poisson probabilities. The
profile-likelihood interval inverts $q_\poi=-2\log\lambda(\poi)$, where
$\lambda(\poi)=L(\poi,\widehat{\nuis}_{\poi})/
L(\widehat{\poi},\widehat{\nuis})$, with threshold
$\chi^2_{1,1-\alpha}$,
while the Bayesian upper limit is the
$(1-\alpha)$ posterior quantile under a flat prior on $\poi\geq0$.

\paragraph*{Spectral calibration.}
The baseline spectral profile-likelihood construction uses
$\chi^2_{2,1-\alpha}$ for joint $(\poi,\mass)$ regions and
$\chi^2_{1,1-\alpha}$ for profiled $\poi$ intervals. The constrained-nuisance
and signed-efficiency studies below instead use their stated toy calibrations.
For LF2I we use the analytic binned likelihood-ratio statistic
$\tau(\mathcal D;\vartheta)=-2[\log L_{\mathcal D}(\vartheta)-
\log L_{\mathcal D}(\widehat\vartheta)]$ for a dataset $\mathcal D$.
Independent nominal simulations at each grid point set its empirical
$(1-\alpha)$ critical value, and inversion retains the points below that
threshold. For the split-conformal construction, the analytic grid posterior
defines the HPD score $S(\vartheta,\mathcal D)$ as the total posterior mass of grid
points whose posterior density exceeds $p(\vartheta\mid\mathcal D)$. From
$n_{\rm cal}$ exchangeable calibration pairs we use the $k$th ordered score, with
$k=\lceil(n_{\rm cal}+1)(1-\alpha)\rceil$, as the threshold and define
$S_{(n_{\rm cal}+1)}=+\infty$. The analytic
statistics are fixed before calibration; no learned statistic is used in the
reported results.

\paragraph*{Coverage estimation.}
At fixed $\vartheta$, effective coverage~\eqref{eq:effective} is estimated
from $N$ nominal or misspecified datasets as the fraction of regions
containing $\theta$, or of projected intervals containing $\poi$, with
binomial standard error $\sqrt{c(1-c)/N}$; for example, $N\approx3600$ gives
$0.005$ near $c=0.9$. Conditional coverage is scanned over $\poi$, including
the boundary, and over $(\poi,\mass)$ in the spectral study; for marginal
coverage, each trial draws the true parameter from the stated prior. Finite
conformal calibration adds a second source of variation: for continuous
scores, coverage conditional on a realised calibration set has a beta
order-statistic variance, which for the discrete Poisson/grid scores used
here is only a reference scale. Where repeated calibrations are available, we
report their empirical spread.

For the projected interval
$\mathcal C_\alpha^\poi(x)=[L(x),U(x)]$, whole-interval coverage can conceal
which endpoint fails. We therefore report the endpoint quantities used above:
\begin{equation}
\begin{gathered}
  c_L(\poi)=\Prob[L(X)\leq\poi],\qquad
  c_U(\poi)=\Prob[U(X)\geq\poi],\\
  c(\poi)=\Prob[L(X)\leq\poi\leq U(X)] .
\end{gathered}
  \label{eq:edge-coverage}
\end{equation}
Here $c_L$ is lower-endpoint (discovery-side) coverage and $c_U$ is
upper-endpoint (exclusion-side) coverage. For a pure upper limit $[0,U]$,
$c_L(0)=c_U(0)=1$ trivially; consequently, evaluating at the boundary does
not test the claimed upper limit.

\paragraph*{Goodness-of-fit rejection probability.}
For each spectral deformation, we fit the nominal model and compute the
saturated Poisson deviance
\begin{equation}
 D_{\rm Pois}=2\sum_i\left[n_i\log\!\left(\frac{n_i}{\hat\nu_i}\right)
 -(n_i-\hat\nu_i)\right],
 \label{eq:gof-deviance}
\end{equation}
where $n_i\log n_i$ is set to zero when $n_i=0$ and $\hat\nu_i$ is the
best-fit nominal spectrum. Independent nominal pseudo-data set the critical
value at type-I error $\alpha_{\rm GoF}=0.05$, and the rejection probability
is the fraction of deformed pseudo-datasets above it: power against the
specified deformation, not signal-discovery power. It is common to all
methods, which use the same binned data. Unless stated otherwise, thresholds
and rejection probabilities each use $4000$ independent spectra, fitted on the
interval-construction grid. For the two-dimensional robustness studies of
Figure~\ref{fig:spectral-robustness}, coverage and
rejection probability are averaged over the same discrete uniform grid prior:
the threshold is the $95$th percentile of the nominal prior-predictive
distribution ($\vartheta\sim\pi_{\rm grid}$,
$X\sim p_0(\cdot\mid\vartheta)$), so the stated $5\%$ type-I error is marginal
over that prior, not uniform point-wise control.

\paragraph*{From omitted deformation to constrained nuisance.}
Misspecification and a modelled systematic uncertainty are distinct statistical
problems. To connect them, we complement the omitted-deformation
experiments with the exactly degenerate model
\begin{equation}
  n\sim\Poisson\!\left[b+\kappa(\poi+\eta)\right],
  \qquad
  a\sim\mathcal N(\eta,\sigma_\eta^2),
  \label{eq:constrained-deformation}
\end{equation}
where the signed nuisance amplitude $\eta$ is expressed in units of signal
strength and $a$ is an independent calibration measurement. The main count
identifies only $\poi+\eta$; hence the auxiliary term, not a goodness-of-fit
test on $n$, identifies $\poi$. We profile $\eta$ in the likelihood ratio
defined above, with $\nuis=\eta$. At every tested $\poi$, the critical value is
the largest toy quantile over a five-point calibration grid spanning
$\eta\in[-0.20,0.20]$, using $8000$ toys per grid point and the $90.5\%$
empirical quantile as a $0.5$ percentage-point guard for the nominal $90\%$
interval. Coverage is evaluated with $20000$ independent toys at
$\eta\in\{-0.20,0,+0.20\}$; this is neither a continuous-range guarantee nor
an extrapolation beyond the calibration grid.

\paragraph*{Signed spectral efficiency stress test.}
At the interior benchmark $(\poi,\mass)=(1,48~\mathrm{GeV})$, the nominal
model expects $30.13$ signal and $40.00$ background events. The uniform loss
lowers the efficiency plateau by up to $50\%$; the threshold shift of
Section~\ref{sec:misspec} preferentially removes low-energy signal events
while the analysis retains the nominal response. We profile over
$\poi\in[0,3]$ and $\mass\in[21,111]~\mathrm{GeV}$, target $95\%$
nominal-model upper-endpoint coverage, and toy-calibrate the one-sided
likelihood-ratio threshold with $10000$ nominal spectra; each severity is
evaluated with $5000$ independent spectra.

\section{Results}
\label{sec:results}

The results test signed endpoint damage, residual detectability, and nuisance
absorption using the coordinates of Section~\ref{sec:results-index}. Each
subsection opens with its main finding, and Table~\ref{tab:summary} collects
them.

\subsection{Exact counting benchmark and its diagnostic limit}
\label{sec:results-l0}

We use the single-bin experiment only where its simplicity is informative:
as an exact check of the interval implementations and as the limiting case in
which a signal-like deformation has $\gamma=0$. The benchmark uses
$\kappa=12$, $b=3$, and $\alpha=0.10$. In a separate $b=0$ implementation
check expressed in signal-count units ($\kappa=1$), the Feldman--Cousins upper
endpoints are the canonical $2.44$, $4.36$, and $5.91$ for $n=0,1,2$.

\subsection{Positive excess at the boundary}
\label{sec:results-robust}

\begin{figure}[t]
  \centering
  \includegraphics[width=\linewidth]{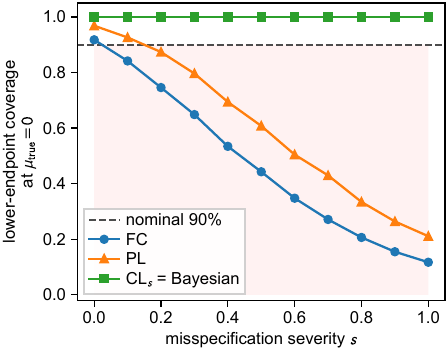}
  \caption{\textbf{Single-bin response to an unmodelled background excess}
  at true $\poi=0$ and $90\%$ confidence. Curves show lower-endpoint
  coverage, which at $\poi=0$ equals whole-interval coverage and is
  identically one for the pure upper limits, versus the misspecification
  severity $s$; at $s=1$ the excess is six expected events, twice the
  nominal background. FC: Feldman--Cousins; PL: profile likelihood;
  the $\mathrm{CL}_s$ and flat-prior Bayesian upper limits coincide. The
  shaded region denotes coverage below $0.90$.}
  \label{fig:robustness}
\end{figure}

A positive, signal-like excess ($\beta>0$, $\gamma=0$) drives lower-endpoint
coverage at the boundary far below nominal, and a single-bin fit cannot
detect it. We inject the background tail of Section~\ref{sec:misspec},
construct each method's interval under the nominal model, and measure coverage at
$\poi=0$, where a unified interval can fail only through its lower endpoint,
as $s$ grows (Figure~\ref{fig:robustness}).

Feldman--Cousins coverage reaches $0.12$ and profile-likelihood coverage
$0.21$ at $s=1$, whereas a pure upper limit $[0,\poi_{\rm up}]$ contains the
boundary by construction and is only weakened. The resolved spectrum breaks
the exact one-bin degeneracy only when the deformation has a component
outside the fitted model (Section~\ref{sec:results-l1robust}).

\subsection{Endpoint-specific coverage away from the boundary}
\label{sec:results-edge}

Away from the boundary, the sign of $\beta$ decides which endpoint fails, and
an upper-limit failure appears only when coverage is evaluated at nonzero
$\poi$. To test the exclusion side directly, we repeat the exactly solvable
counting experiment at $\poi_{\rm true}=0.25$. This benchmark gives three nominal
signal events for $\kappa=12$, equal to the nominal background $b=3$, and is
a signal at the scale of the experiment's sensitivity: the median expected
$90\%$ Feldman--Cousins upper limit under the background-only hypothesis is
$\poi\simeq0.37$. Alongside the background excess, which biases the signal
upward, we apply the signal-efficiency overestimate of
Section~\ref{sec:misspec}, which biases it downward, and compute the three
probabilities of Eq.~\eqref{eq:edge-coverage} exactly.
At this reference point $\nu_0=\kappa\poi+b=6$. Because a one-bin
deformation is exactly signal-collinear when $b$ is fixed,
$\gamma=0$ and $\beta=\delta\nu/\sqrt{\nu_0}$. The maximal background
excess therefore has $(\beta,\gamma)=(+2.45,0)$, whereas the $80\%$
signal-efficiency loss has $(\beta,\gamma)=(-0.98,0)$.

\begin{figure*}[t]
  \centering
  \includegraphics[width=\textwidth]{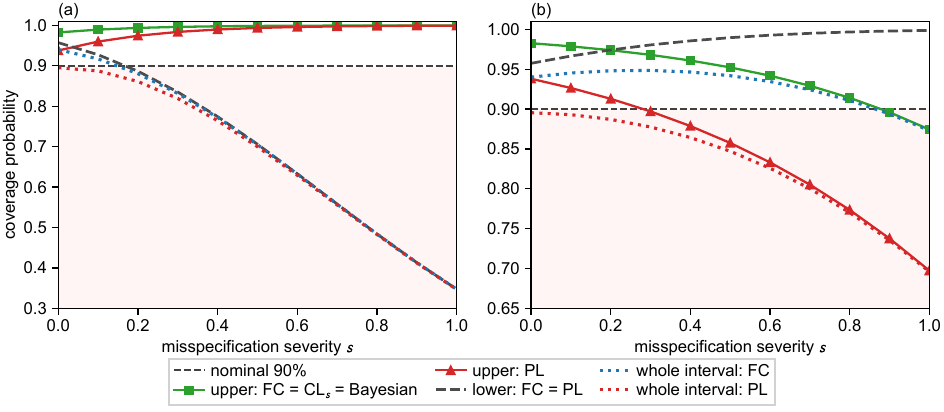}
  \caption{\textbf{Endpoint-specific coverage under signed model
  distortions} in the counting experiment
  ($\poi_{\rm true}=0.25$, three nominal signal events, $90\%$ CL).
  Solid curves show upper-endpoint coverage, dashed curves show
  lower-endpoint coverage, and dotted curves show whole-interval coverage
  for Feldman--Cousins (FC) and profile likelihood (PL). (a) Up to six
  unmodelled background events are added; (b) the true signal efficiency is
  reduced by up to $80\%$. All probabilities are exact Poisson sums.
  Coinciding curves share a legend entry, marked ``='': FC,
  $\mathrm{CL}_s$, and the flat-prior Bayesian upper limit all have
  $U(n)<\poi_{\rm true}$ only for $n\leq1$, and the FC and PL
  lower-endpoint coverage curves coincide at this benchmark because
  $L(n)\leq\poi_{\rm true}$ selects the same counts for both methods.}
  \label{fig:edge-coverage}
\end{figure*}

The direction of the effects in Figure~\ref{fig:edge-coverage} is the one
expected from the sign of $\beta$ in Eq.~\eqref{eq:local-endpoint-coverage}:
a larger (smaller) true mean shifts the interval up (down). At zero
severity, upper-endpoint coverage
is $0.983$ for Feldman--Cousins and both pure upper limits, and $0.938$ for
the profile likelihood; the corresponding whole-interval coverages are
$0.940$ and $0.895$ for Feldman--Cousins and profile likelihood. Under the
maximal positive background excess, upper-endpoint coverage tends to one
for every method, while lower-endpoint and whole-interval coverage both fall
to $0.347$ for Feldman--Cousins and profile likelihood. Thus the same
misspecification that creates false discoveries makes upper limits more
conservative, not anti-conservative.

Reversing the sign of the deformation reverses the threatened endpoint. At
the $80\%$ signal-efficiency loss stress-test endpoint, lower-endpoint
coverage is essentially one, but upper-endpoint coverage falls to $0.874$ for
Feldman--Cousins, $\mathrm{CL}_s$, and the Bayesian upper limit, and to
$0.697$ for the profile likelihood; whole-interval coverage follows the upper
endpoint ($0.873$ and $0.696$). Such differences between constructions at a
single $\poi_{\rm true}$ largely reflect the discreteness of the Poisson
belts: the Feldman--Cousins upper endpoint for $n=2$ is $0.253$, just above
$\poi_{\rm true}$, and at $\poi_{\rm true}=0.255$ its upper-endpoint coverage is already
$0.699$, as for the profile likelihood. The sign dependence is not specific
to this benchmark: because the loss removes a fixed fraction of the signal,
$|\beta|$ grows with $\poi_{\rm true}$, and at $\poi_{\rm true}=0.5$
($\beta=-1.60$) upper-endpoint coverage falls from $0.945$ to $0.410$ for all
four constructions. The $80\%$ loss is a controlled endpoint chosen to make
the sign dependence visible, not a representative efficiency uncertainty.
Claiming a robust limit therefore requires upper-endpoint coverage at nonzero
$\poi$ and for both deformation signs; the $\poi=0$ check of
Section~\ref{sec:results-robust} alone cannot establish it.

\subsection{Promoting a signal-like deformation to a constrained nuisance}
\label{sec:results-nuisance}

For a silent, signal-like deformation, modelling it as a constrained nuisance
(step~4 of Section~\ref{sec:results-index}) restores coverage at a
measurable cost in interval length. We repeat the signed stress test after
replacing the omitted deformation by the nuisance-aware likelihood of
Eq.~\eqref{eq:constrained-deformation}.
The benchmark remains $\poi_{\rm true}=0.25$, $\kappa=12$, and $b=3$; the
endpoints $\eta_{\rm true}=\mp0.20$ shift the main-count mean by $\mp2.4$
events, and $\sigma_\eta$ is varied from $0.01$ to $0.40$. The
omitted-deformation analysis fixes $\eta=0$ and uses the exact
Feldman--Cousins belt.

\begin{figure*}[t]
  \centering
  \includegraphics[width=\textwidth]{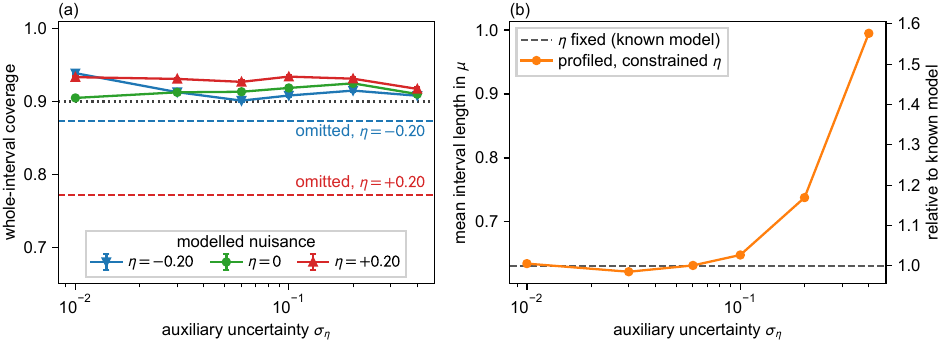}
  \caption{\textbf{Coverage--sensitivity trade-off for a constrained
  signal-like deformation} at $\poi_{\rm true}=0.25$ and $90\%$ CL.
  (a) Whole-interval coverage after profiling the signed nuisance
  $\eta$ using a five-point calibration grid on $[-0.20,0.20]$ and evaluating
  $\eta_{\rm true}\in\{-0.20,0,+0.20\}$; error bars use $20000$
  pseudo-experiments. Dashed lines omit the deformation; the dotted line marks
  nominal $90\%$ coverage. (b) Mean interval length at $\eta_{\rm true}=0$
  versus auxiliary uncertainty; the dashed line is the known-model reference
  with $\eta$ fixed, and the right-hand axis gives the length relative to it.}
  \label{fig:nuisance-aware}
\end{figure*}

Figure~\ref{fig:nuisance-aware} separates the cost of acknowledging an
uncertainty from the damage caused by omitting it. When $\eta$ is fixed
incorrectly to zero, whole-interval coverage is $0.873$ at
$\eta_{\rm true}=-0.20$ and $0.772$ at $\eta_{\rm true}=+0.20$, failing
through the upper and the lower endpoint, respectively. After $\eta$ is
included and constrained, the measured
whole-interval coverage across all six auxiliary precisions is
$0.901$--$0.939$ for the negative deformation, $0.905$--$0.925$ at
$\eta=0$, and $0.917$--$0.934$ for the positive deformation.

At the evaluated points, coverage restoration is not free. At the nominal
point $\eta=0$, the mean interval length grows from $0.63$ in the known-model
reference to $1.0$ at $\sigma_\eta=0.40$, a factor of $1.6$. In the formal
limit $\sigma_\eta\to\infty$, the likelihood
depends on $\poi$ and $\eta$ only through their sum, so $\poi$ is not
identifiable and no finite data-driven interval exists without an additional
constraint or a bounded conservative envelope.

\subsection{Nominal spectral calibration}
\label{sec:results-l1}
\label{sec:results-sbi}

Before any deformation, the calibrated spectral constructions meet their
stated targets, LF2I point-wise and conformal-NPE marginally, so the coverage
losses below are due to misspecification. Analytic binned likelihood ratios
and grid posteriors isolate calibration from estimator approximation. In a
$29$-point signal-strength audit at $\mass=\SI{48}{GeV}$, with $40000$
evaluation and $40000$ LF2I calibration spectra per point and $40000$
conformal calibration pairs, LF2I joint-region coverage lies between $0.8966$
and $0.9034$. Across 20 conformal calibrations, marginal joint-region
coverage has mean $0.8997$ (range $0.8975$--$0.9021$), while point-wise
coverage is non-uniform and reaches $0.8624$.

\subsection{Low-energy tail in the spectral analysis}
\label{sec:results-l1robust}

\begin{figure*}[t]
  \centering
  \includegraphics[width=\textwidth]{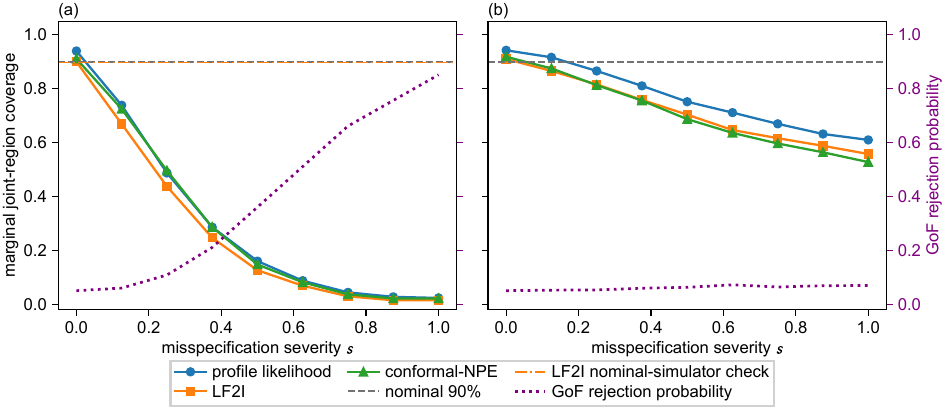}
  \caption{\textbf{Spectral robustness to background and halo deformations.}
  (a) Unmodelled low-energy tail, adding up to 40 expected events at
  $s=1$. (b) Halo-model shift defined in Section~\ref{sec:misspec}.
  In both panels, marginal joint-region coverage (solid curves, left scale)
  and goodness-of-fit rejection probability (purple dotted curve, right scale)
  are averaged over the same discrete uniform $(\poi,\mass)$ grid prior.
  The GoF threshold has marginal $5\%$ type-I error. Dashed lines mark nominal
  $90\%$ coverage; the orange dash-dotted line in (a) is LF2I's
  nominal-simulator coverage check. Coverage uses $1500$ spectra per
  severity in (a) and $3000$ in (b); GoF probabilities use $4000$.}
  \label{fig:spectral-robustness}
\end{figure*}

A low-energy tail has a component outside the fitted WIMP family, so a
goodness-of-fit test can detect it, but only once it is large. Each method
uses the nominal spectral model and is evaluated as the tail
grows, with joint-region coverage and deviance rejection probability averaged
over the same discrete uniform $(\poi,\mass)$ grid prior
(Figure~\ref{fig:spectral-robustness}a).

All three methods lose coverage. LF2I's internal diagnostic remains nominal
because it uses fresh nominal-simulator draws, whereas the data-based
rejection probability rises from $0.05$ to $0.85$; this comparison is
marginal over the grid prior and does not imply uniform point-wise behaviour.
At the boundary point $(\poi,\mass)=(0,\SI{48}{GeV})$, a quarter of the
maximal tail (10 events) already drives conditional profile-likelihood
coverage to $0.01$, while the rejection probability is still $0.05$: its
coordinates $(\beta,\gamma)=(7.25,0.55)$ place it in the silent corner
(Table~\ref{tab:summary}).

\subsection{Energy-dependent efficiency loss and the full fitted tangent space}
\label{sec:results-spectral-efficiency}

\begin{figure*}[t]
  \centering
  \includegraphics[width=\textwidth]{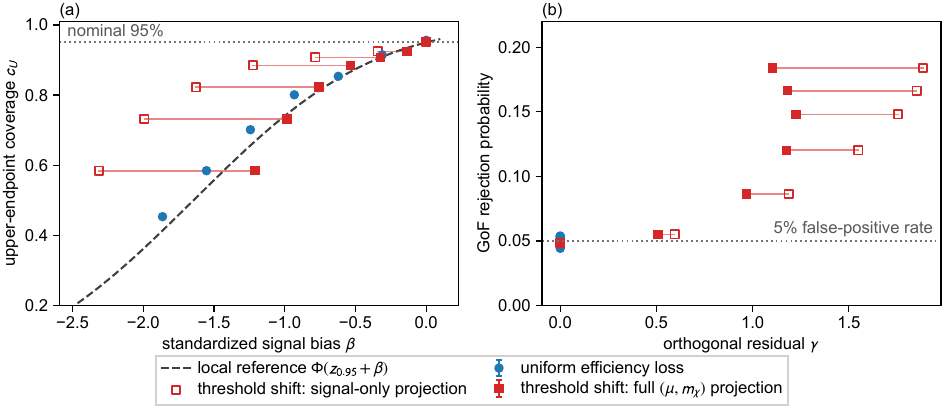}
  \caption{\textbf{Signed spectral efficiency stress test.}
  (a) Toy-calibrated upper-endpoint coverage versus the signed
  standardized bias, with the local reference (dashed). (b) Goodness-of-fit
  rejection probability versus the orthogonal residual. Filled points use the
  full $(\poi,\mass)$ tangent space; open squares show the signal-only
  projection of the threshold-loss deformation. Horizontal connectors join
  alternative projections of the same measured point.}
  \label{fig:spectral-efficiency}
\end{figure*}

A negative detector-response error threatens the upper endpoint; accounting
for all fitted directions substantially improves the local prediction of
the coverage loss in the threshold-shift study. At the nonzero-signal
benchmark of Section~\ref{sec:protocol},
Figure~\ref{fig:spectral-efficiency} compares the measured upper-endpoint
coverage and goodness-of-fit rejection probability with the signed Fisher
projection of Eq.~\eqref{eq:index}. For the threshold loss, the projection is
computed both onto the signal-normalization direction alone and onto the full
local $(\poi,\mass)$ tangent space, two predictions for the same pseudo-data.

The uniform loss provides the exact collinear control. Its orthogonal
residual is $\gamma=0$ at every severity, and the goodness-of-fit rejection
probability remains between $0.044$ and $0.054$, consistent with the $5\%$
false-positive floor. At the $50\%$ loss endpoint, the full-space
standardized bias is $\beta=-1.86$ and upper-endpoint coverage falls from
$0.956$ to $0.453$. The signed relation
$c_U\simeq\Phi(z_{0.95}+\beta)$ predicts $0.413$. The four-percentage-point
discrepancy exceeds the Monte Carlo standard error of $0.007$ and illustrates
the quantitative limitations of the local Gaussian approximation in this
low-count, finite-grid study.

The energy-dependent case demonstrates why the full fitted space matters.
At the maximal threshold shift, the integrated signal yield is reduced by
$44\%$. A signal-only calculation assigns
$(\beta,\gamma)=(-2.31,1.88)$ to this deformation. Once the mass direction
is included, $86.4\%$ of its squared Fisher norm is absorbed by the fitted
tangent space, giving
$(\beta,\gamma)=(-1.21,1.10)$ and predicted local parameter shifts
$(\Delta\hat\poi,\Delta\hat\mass)=(-0.32,+34.6~\mathrm{GeV})$. The measured
upper-endpoint coverage is $0.584$, much closer to the full-space prediction
$0.669$ than to the signal-only prediction $0.252$, although the remaining
$0.085$ discrepancy exceeds the Monte Carlo standard error of $0.007$.
Meanwhile the
goodness-of-fit rejection probability reaches only $0.184$. Thus the fitted
mass both shields part of the upper endpoint from the larger signal-only
bias and hides part of the spectral response error from the diagnostic; a
single-template calculation overstates both damage and detectability.

\subsection{Halo-model misspecification}
\label{sec:results-halo}

A halo shift is absorbed almost entirely by the fitted WIMP mass:
joint-region coverage falls while the goodness-of-fit test has little power
above its false-positive rate. Varying the astrophysical Standard Halo Model changes
the accepted WIMP spectrum through both its shape and thresholded yield. For
the joint halo-parameter shift of Section~\ref{sec:misspec}
(Figure~\ref{fig:spectral-robustness}b), marginal joint-region coverage at the
stress-test endpoint is $0.61$ for the profile likelihood and $0.56$ and
$0.53$ for LF2I and conformal-NPE, while the goodness-of-fit rejection
probability reaches only about $0.07$. At the interior reference point the
signal-strength coordinate is small, $\beta_\poi=0.106$, whereas the local
mass shift is $18.1~\mathrm{GeV}$: the example shows low diagnostic power
against a fitted secondary-parameter displacement, not a large
signal-strength bias or an upper-endpoint failure.

\subsection{A second domain: the \texorpdfstring{$\nubb$}{0nu-beta-beta} peak}
\label{sec:results-nubb}

\begin{figure*}[t]
  \centering
  \includegraphics[width=\textwidth]{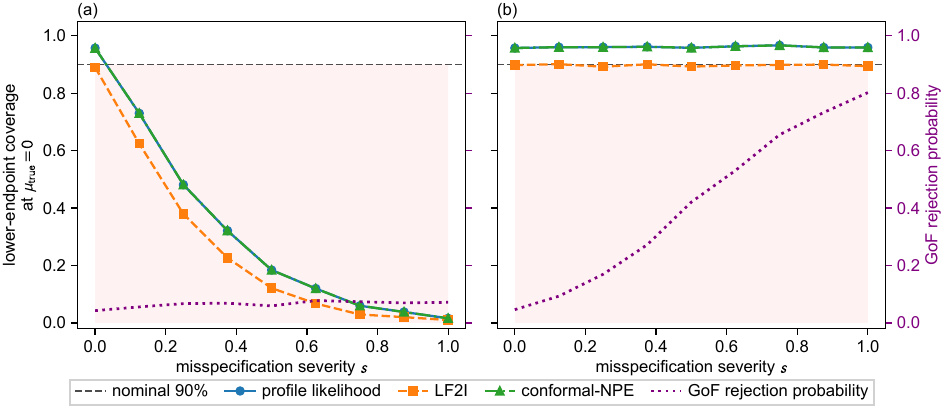}
  \caption{\textbf{$\nubb$ line deformation.} Conditional whole-interval
  coverage at $\poi_{\rm true}=0$ --- equivalently lower-endpoint coverage ---
  versus the amplitude of an unmodelled $\gamma$ line, centred at the
  $Q$-value (a) or displaced by $\SI{30}{keV}$ (b). Purple dotted curves
  (right scale) are rejection probabilities of the $5\%$ saturated-Poisson
  goodness-of-fit test; dashed lines mark nominal $90\%$ coverage. Each
  severity uses $3000$ spectra for coverage and $2000$ for GoF; the GoF
  threshold uses $2000$ independent nominal spectra.}
  \label{fig:nubb}
\end{figure*}

In a second physics domain, the same unmodelled $\gamma$ line is silent and
damaging at the $Q$-value but detectable and harmless when displaced. We
consider a $^{76}\mathrm{Ge}$ $\nubb$ peak search
with $Q_{\beta\beta}=\SI{2039}{keV}$, $\sigma_E=\SI{1.2}{keV}$, four signal
counts at $\poi=1$, and eight continuum counts in a
$\pm\SI{50}{keV}$ window, analysed with the same three spectral
constructions. Localized contamination is closely related to the ``spurious
signal'' tests used to screen background functions in diphoton resonance
analyses~\cite{ATLAS2014Spurious,CMS2014Diphoton}. The deformation is a weak
$\gamma$-ray line with six expected counts at $s=1$, and its position
determines the result (Figure~\ref{fig:nubb}). At the $Q$-value it is
indistinguishable from signal: lower-endpoint coverage falls to
$0.01$--$0.02$, while the goodness-of-fit rejection probability remains below
$0.08$. Displaced by $\SI{30}{keV}$ ($25\,\sigma_E$), the same line leaves
lower-endpoint coverage at its nominal-model value and raises the rejection
probability from $0.05$ to $0.80$. Spectral resolution thus helps only when
the deformation is identifiable against the full fitted model.

\section{Empirical assessment of the degeneracy coordinates}
\label{sec:results-index-validation}

\begin{figure*}[t]
  \centering
  \includegraphics[width=\textwidth]{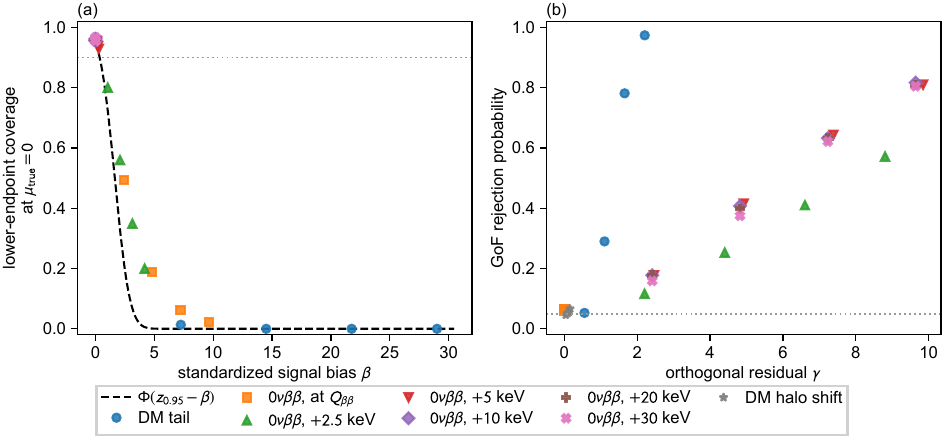}
  \caption{\textbf{Empirical assessment of the local degeneracy
  coordinates.} (a) Conditional profile-likelihood coverage at true
  $\poi=0$ versus $\beta$ for the dark-matter tail and $\nubb$ line offsets;
  the dashed curve is the local Gaussian reference
  $\Phi(z_{0.95}-\beta)$. (b) Rejection probability of the $5\%$
  goodness-of-fit test versus $\gamma$. Symbols identify the deformations,
  one point per severity; dotted lines mark nominal $90\%$ coverage (a) and
  the $5\%$ false-positive rate (b). Coverage uses $3000$ spectra per
  point; GoF thresholds and probabilities each use $4000$.}
  \label{fig:index}
\end{figure*}

Table~\ref{tab:summary} collects representative configurations of the
deformation families, with their index coordinates, measured coverage, and
goodness-of-fit rejection probability. Figure~\ref{fig:index} shows the
boundary cases as functions of $\beta$ and $\gamma$. The coordinates help
screen endpoint vulnerability and residual detectability, but neither gives
a universal ordering across deformations or a finite-sample calibration.

\begin{table*}[t]
\caption{Index coordinates and measured operating characteristics of the
studied deformations, at the stress-test endpoint unless stated. Coverage is
that of the profile-likelihood construction for the listed endpoint of a
nominal $90\%$ interval; individual endpoint coverages need not equal $0.90$.
The spectral efficiency studies instead target $95\%$ upper-endpoint
coverage. The halo row gives marginal joint-region coverage of the three
spectral constructions; its index is evaluated at the stated reference point.
Tail and line rows use the independent ensembles of Figure~\ref{fig:index}.
GoF: rejection
probability of the $5\%$ goodness-of-fit test, not evaluated for the counting studies.
Reference points: $\poi=0$ (tail, lines), $(\poi,\mass)=(1,\SI{48}{GeV})$
(spectral efficiency losses), $(1.05,\SI{48}{GeV})$ (halo). \emph{Silent}: coverage
loss with the goodness-of-fit test near its false-positive rate, the case
that step~4 of Section~\ref{sec:results-index} promotes to a constrained
nuisance.}
\label{tab:summary}
\begin{tabular*}{\textwidth}{@{\extracolsep\fill}lrrlccl@{}}
\toprule
Deformation & $\beta$ & $\gamma$ & Endpoint & Coverage & GoF & Regime\\
\midrule
\multicolumn{7}{@{}l}{\emph{Counting experiment, $\poi_{\rm true}=0.25$
 (Section~\ref{sec:results-edge})}}\\
Background excess, 6 events & $+2.45$ & $0$ & lower & $0.347$ & -- & silent\\
Signal-efficiency loss, $80\%$ & $-0.98$ & $0$ & upper & $0.697$ & -- & silent\\
\midrule
\multicolumn{7}{@{}l}{\emph{Dark-matter spectrum
 (Sections~\ref{sec:results-l1robust}--\ref{sec:results-halo})}}\\
Low-energy tail, 10 events & $+7.25$ & $0.55$ & lower & $0.01$ & $0.05$ & silent\\
Low-energy tail, 40 events & $+29.0$ & $2.21$ & lower & $0.00$ & $0.97$ & detected\\
Uniform efficiency loss, $50\%$ & $-1.86$ & $0$ & upper & $0.453$ & $0.05$ & silent\\
Threshold shift, 3$\to$\SI{9}{keV} & $-1.21$ & $1.10$ & upper & $0.584$ & $0.18$ & weakly detected\\
Halo shift & $+0.11$ & $0.16$ & joint & $0.53$--$0.61$ & $0.07$ & silent\\
\midrule
\multicolumn{7}{@{}l}{\emph{$\nubb$ peak, line of 6 events
 (Section~\ref{sec:results-nubb})}}\\
Line at $Q_{\beta\beta}$ & $+9.65$ & $0$ & lower & $0.02$ & $0.06$ & silent\\
Line at $Q_{\beta\beta}+\SI{2.5}{keV}$ & $+4.18$ & $8.81$ & lower & $0.20$ & $0.57$ & partly detected\\
Line at $Q_{\beta\beta}+\SI{30}{keV}$ & $0.00$ & $9.65$ & none & $0.96$ & $0.80$ & detected, unbiased\\
\botrule
\end{tabular*}
\end{table*}

In Figure~\ref{fig:index}, the tail and line points are evaluated at
$\poi=0$, where the local reference is $c_L\simeq\Phi(z_{1-\alpha/2}-\beta)$,
and the halo residual at $(\poi,\mass)=(1.05,\SI{48}{GeV})$; the negative
efficiency losses appear in
Figures~\ref{fig:edge-coverage} and~\ref{fig:spectral-efficiency}. Three
features stand out.

First, the Gaussian formula is not a quantitative finite-sample calibration
at large deformation: at $\beta=4.2$ it gives approximately $0.005$, whereas the
measured coverage is $0.20$. We therefore use $\beta$ for screening and rely
on toy calibration for numerical coverage statements in the low-count regime.

Second, small $\gamma$ is associated with low goodness-of-fit power in the
studied cases, even when coverage is badly degraded: the $\nubb$ line at the $Q$-value
($\gamma=0$) and the maximal halo shift, $97.1\%$ of whose squared Fisher norm
is absorbed by the local
$(\partial\nu/\partial\poi,\partial\nu/\partial\mass)$ tangent, are rejected
with probabilities $0.06$ and $0.07$. Within the dark-matter analysis, the
rejection probability for the tail rises from $0.05$ to $0.97$ as $\gamma$
grows with the tail amplitude from $0.55$ to $2.21$. This trend is not a
universal power curve: the tail at $\gamma=2.21$ has rejection probability
$0.97$, whereas the displaced $\nubb$ line at $\gamma=8.81$ has probability
$0.57$. Even within the threshold-shift study, the residual norm decreases
at the largest severities while measured power increases. Such nonlocal,
low-count behaviour requires toy calibration of the specified test.

Third, varying the $\nubb$ line offset resolves the \emph{orthogonal
transition}: at $\SI{2.5}{keV}$ ($\approx 2\,\sigma_E$) the deformation splits
between the two axes, and by $\SI{5}{keV}$ ($\approx 4\,\sigma_E$) the
orthogonal component dominates ($\beta=0.29$, $\gamma=9.85$), with
lower-endpoint coverage back to $0.93$ and rejection probability $0.81$.

\section{Discussion}
\label{sec:discussion}

Our results clarify four principles for reporting low-background limits,
followed by the study's limitations.

\textbf{Upper-limit coverage and rejection of the boundary are distinct
claims.} Whole-interval coverage at $\poi=0$ tests whether a lower endpoint
spuriously excludes the no-signal hypothesis; it cannot validate a pure
upper limit $[0,\poi_{\rm up}]$, which contains the boundary by construction.
Robustness studies should therefore report $c_L$ and $c_U$ separately, at
nonzero $\poi$ and for both deformation signs. Energy-dependent losses must
also be projected against all fitted directions, including secondary signal
parameters. For example, in the threshold-shift study of
Section~\ref{sec:results-spectral-efficiency}, profiling the WIMP mass
changes $(\beta,\gamma)$ from $(-2.31,1.88)$ to $(-1.21,1.10)$, reducing both
the magnitude of
the predicted bias and the residual available to a goodness-of-fit test.

\textbf{Nominal guarantees are model-specific.} Each method must be reported
against its actual target, point-wise or marginal
(Section~\ref{sec:protocol}), and none is automatically protected against
misspecification: LF2I, conformal-NPE, and profile likelihood all lose
joint-region coverage under the studied spectral deformations,
while LF2I's internal check remains nominal because it uses nominal draws.
Simulation-based guarantees remain conditional on simulator fidelity.

\textbf{Interval methods require a separate model check.} The saturated
single-bin fit has no residual sensitivity to a positive signal-collinear
excess, and spectra add diagnostic power only for components orthogonal to
the complete fitted family. Intervals should therefore be paired with a
goodness-of-fit test whose binning and range retain sensitivity to plausible
deformations (step~4 of Section~\ref{sec:results-index}).

\textbf{From diagnostic to action.} The procedure of
Section~\ref{sec:results-index} uses $\mathcal I_{\mathrm{PF}}$ to triage
candidate deformations before any toy calibration, along two axes: the signed
shift of the fitted signal strength in profiled standard-error units
($\beta$), and the residual norm that locally indicates goodness-of-fit
detectability ($\gamma$). Table~\ref{tab:summary}
places representative deformation configurations in this plane, and
Figure~\ref{fig:nuisance-aware} quantifies the remedy of step~4 in the
exact-degeneracy limit. Near, rather than exact, collinearity produces weak
identification, which also requires explicit calibration.

\textbf{Limitations.} The studies use one counting bin or a two-parameter
spectrum, analytic likelihood ratios, and grid posteriors; learned and
high-dimensional architectures are outside scope, and grid coverage retains
small off-grid resolution effects. The four controlled deformation families
are stress tests, not empirical uncertainty envelopes. The spectral detector
study varies one efficiency turn-on, rather than correlated calibration,
resolution, and background-response fields; background-efficiency loss is
treated analytically, not numerically. The nuisance extension is exactly
collinear, calibrates on five points, and evaluates three. It therefore does
not establish uniform coverage between grid points or beyond the auxiliary
constraint; high-dimensional nuisance fields require a dedicated study.

\section{Conclusions}
\label{sec:conclusions}

We introduce and assess the Poisson--Fisher degeneracy index
$\mathcal I_{\mathrm{PF}}=(\beta,\gamma)$, a nuisance-aware diagnostic that
maps a specified model deformation to its signed effect on signal inference
and the residual available to a goodness-of-fit test, after accounting for
the complete fitted parameter space. The sign of $\beta$ agrees with the
direction of the endpoint failures studied here, while $\gamma$ helps
identify cases with low residual-test power. Neither coordinate provides a
universal ordering across deformation families, and
Eq.~\eqref{eq:local-endpoint-coverage} is not a finite-sample calibration.
Toy studies remain necessary, especially at boundaries and for nonlocal
deformations. Spectral information helps only when a deformation is
identifiable against the full fitted model.

Model misspecification acts through the sign of the induced signal bias:
positive signal-like
contamination can destroy lower-endpoint coverage while making the upper
endpoint conservative, whereas negative signal bias can make it
anti-conservative. Coverage at $\poi=0$ diagnoses lower-endpoint rejection,
not a pure upper limit. Calibration under the nominal simulator does not
protect against either failure when that simulator is misspecified relative
to the data-generating process.

A plausible large-$|\beta|$, small-$\gamma$ direction should be represented
as a nuisance constrained by auxiliary data or a defensible envelope. In the
exactly collinear benchmark, this removes the observed failures at the three
evaluated nuisance values while increasing the mean interval length by up to
a factor of $1.6$. Without such external information, signal and deformation
are not separately identifiable, and a finite model-dependent limit should
not be presented as robust.

\backmatter

\bibliography{references}

\end{document}